\pdfoutput=1
\documentclass[11pt]{article}
\usepackage{makecell}
\usepackage[final]{acl}

\newcommand{\dq}[1]{``#1''}

\usepackage{times}
\usepackage{latexsym}
\usepackage{amsmath}
\usepackage{placeins}
\usepackage{booktabs} 
 \usepackage{booktabs}
 \usepackage{siunitx}
 \usepackage{hyperref}
 \usepackage{float}     %
 \usepackage{placeins}  %

\usepackage[T1]{fontenc}

\usepackage[utf8]{inputenc}

\usepackage{microtype}

\usepackage{inconsolata}

\usepackage{graphicx}
\usepackage{booktabs}
\usepackage{siunitx}   %
\usepackage{hyperref}  %
\usepackage{float}     %
\usepackage{cuted}     %
\usepackage{capt-of}   %

\title{NER Retriever: Zero-Shot Named Entity Retrieval \\ with Type-Aware Embeddings}

\author{\makecell{Or Shachar~~~~~Uri Katz~~~~~Yoav Goldberg~~~~~Oren Glickman}\\ Computer Science Department, Bar-Ilan University \\
\texttt{\{shachao8, uri.katz, yoav.goldberg, oren.glickman\}@biu.ac.il} 
}

\begin{document}
\maketitle
\begin{abstract}
We present \textbf{NER Retriever}, a zero-shot retrieval framework for ad-hoc Named Entity Retrieval, a variant of Named Entity Recognition (NER), where the types of interest are not provided in advance, and a user-defined type description is used to retrieve documents mentioning entities of that type. Instead of relying on fixed schemas or fine-tuned models, our method builds on internal representations of large language models (LLMs) to embed both entity mentions and user-provided open-ended type descriptions into a shared semantic space. We show that internal representations, specifically the value vectors from mid-layer transformer blocks, encode fine-grained type information more effectively than commonly used top-layer embeddings. To refine these representations, we train a lightweight contrastive projection network that aligns type-compatible entities while separating unrelated types. The resulting entity embeddings are compact, type-aware, and well-suited for nearest-neighbor search. Evaluated on three benchmarks, NER Retriever significantly outperforms both lexical and dense sentence-level retrieval baselines. Our findings provide empirical support for representation selection within LLMs and demonstrate a practical solution for scalable, schema-free entity retrieval. The NER Retriever Codebase is publicly available at: \url{https://github.com/ShacharOr100/ner_retriever}
\end{abstract}

\section{Introduction}
\begin{figure}[!t]
    \centering
\includegraphics[width=0.95\linewidth]{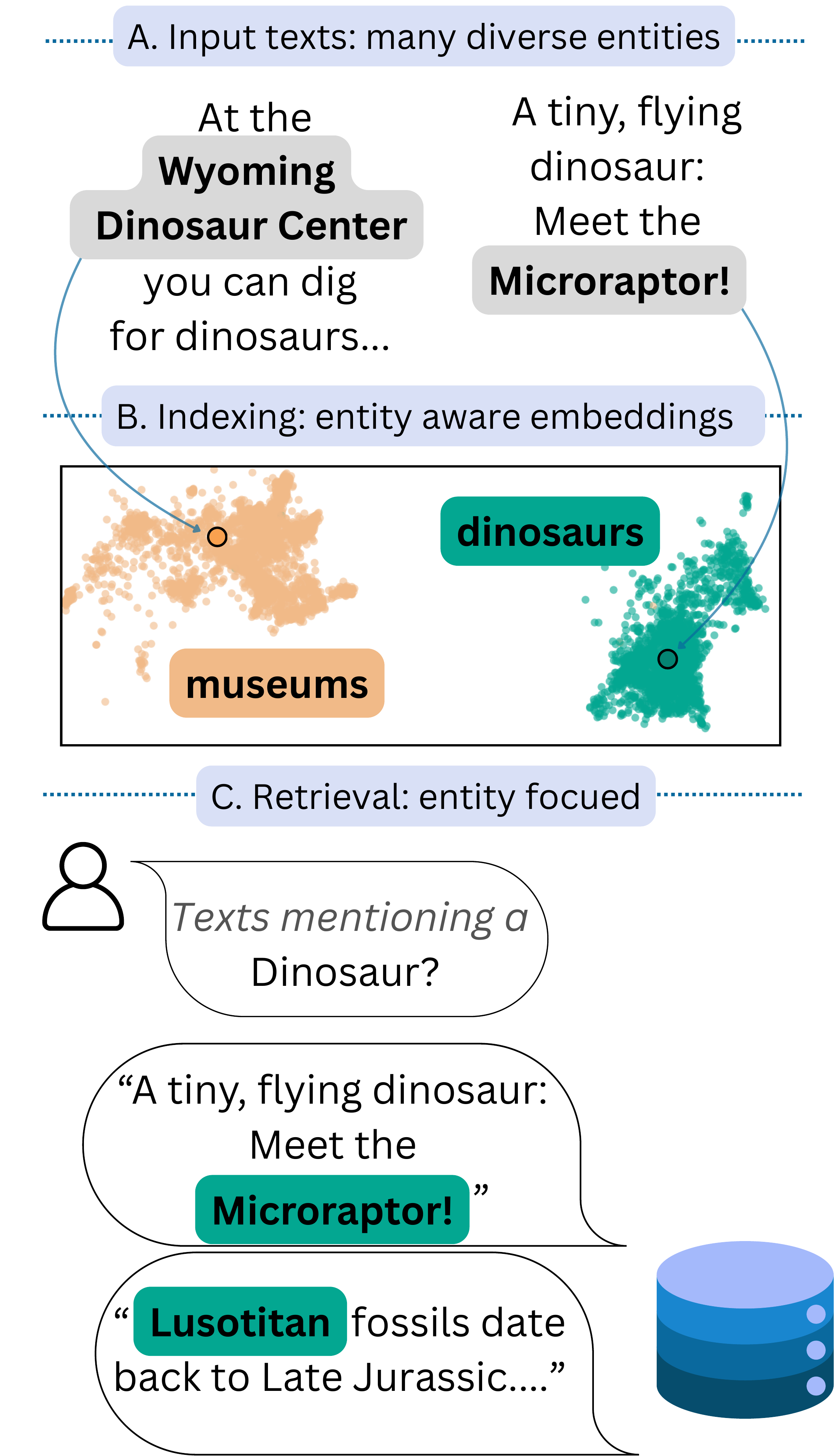}
    \caption{Example use case for ad-hoc Named Entity Retrieval given a query type (“dinosaur”)}
\label{fig:main_illustration}
\end{figure}

Named Entity Recognition (NER) is a foundational task in natural language processing (NLP), aimed at identifying and classifying entity mentions in text into predefined categories. While traditional NER systems have made substantial progress, they typically rely on extensive labeled data and are constrained to a fixed set of entity types such as \textit{person}, \textit{organization} or \textit{location}. This limits their applicability in real-world settings, where the types of entities of interest can be diverse, domain-specific, and defined in an ad-hoc manner.
To address this limitation, fine-grained entity typing \cite{choi-etal-2018-ultra} assigns open-ended fine-grained entity types to entities.
Recently, it was proposed to extend named-entity recognition to named entity \emph{retrieval} \cite{katz-etal-2023-neretrieve}. In the retrieval task
rather than predicting entity types from a closed schema, the goal is to store documents in a way that enables retrieving all text segments that mention entities of a type defined at query time (e.g., ``retrieve all texts mentioning a \textit{dinosaur} or a \textit{politician}''). This setup reflects practical needs in information retrieval, question answering, and knowledge base construction. However, it poses unique challenges: entity types may be unseen during training, and representations must generalize across fine-grained, open-ended categories.

In this work, we propose \textbf{NER Retriever}, a zero-shot retrieval framework for ad-hoc Named Entity Retrieval. Our method leverages the internal structure of large language models (LLMs) to obtain type-aware representations of entity mentions. Specifically, we extract contextual embeddings from an intermediate transformer layer (e.g., layer 17 of LLaMA 3.1 8B) and distill them into a compact, discriminative embedding space using a contrastive learning objective. This allows the system to generalize to novel entity types without requiring task-specific LLM finetuning.

Figure~\ref{fig:main_illustration} illustrates a typical use case. During offline indexing, all entity mentions in the corpus (e.g., ``Microraptor'', ``Lusotitan'') are projected into an embedding space where they cluster near their type description (e.g., ``dinosaur'') and remain distinct from unrelated concepts (e.g., ``The Wyoming Dinosaur Museum''). At query time, user-specified type queries are mapped into the same space and used to retrieve relevant documents via nearest-neighbor search.

We evaluate NER Retriever across three benchmarks: Few-NERD \cite{ding-etal-2021-nerd}, MultiCoNER 2 \cite{fetahu-etal-2023-multiconer}, and NERetrieve \cite{katz-etal-2023-neretrieve}, and find that it outperforms both lexical (BM25) and dense retrieval baselines in most cases. Our results demonstrate that the system enables zero-shot named entity retrieval of highly fine-grained entity types across diverse domains, as represented in the evaluation benchmarks. Our results highlight the value of mid-layer LLM representations and contrastive tuning for zero-shot entity retrieval, and offer insights into the internal structure of type-sensitive knowledge within pretrained language models.

\section{Task Formulation and Problem Definition}
\label{sec:task_formulation}
We follow the task formulation proposed by \citet{katz-etal-2023-neretrieve}, which defines \textit{Named Entity Retrieval} as retrieving texts that mention entities belonging to a given entity type. Unlike traditional NER, which detects and classifies entity spans in text, this task frames the entity type itself as an open-ended user-provided query (e.g., \dq{mountain}, \dq{politician}, \dq{dinosaurs}) and aims to return all documents or sentences containing mentions of entities of that type.
Formally, given a query $q_{\text{type}}$ representing a specific named entity type, and a corpus $\mathcal{C} = \{d_1, d_2, \ldots, d_N\}$ of documents, the task is to return a ranked list $\mathcal{L} = [d_i, d_j, \ldots, d_k]$ of documents from $\mathcal{C}$. Each document in $\mathcal{L}$ must contain at least one named entity mention whose type corresponds to the query $q_{\text{type}}$. The task is evaluated under an \emph{exhaustive} retrieval setting, where systems are expected to return all relevant documents for each query type from a large corpus. The task is designed to support zero-shot retrieval, a setup that reflects real-world scenarios in which users may search for novel or emerging entity categories, and systems must generalize beyond memorized types to retrieve relevant content.

\begin{figure*}[!t]
    \centering
\includegraphics[width=0.99\linewidth]{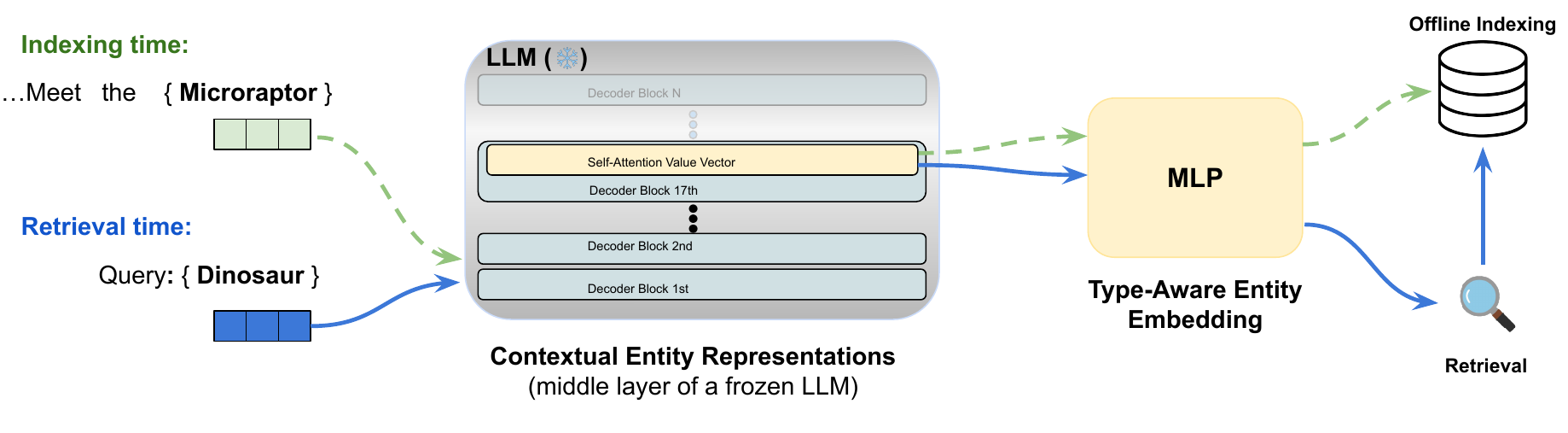}
    \caption{NER Retriever high-level architecture. During indexing (green, dashed), entity spans are embedded and stored using LLM mid-layer contextual representation and a type-aware encoder. At query time (blue, solid), the type description is embedded using the same pipeline and matched to relevant entities via nearest-neighbor search.}
\label{fig:main_architecture}
\end{figure*}

\section{Method Overview}
\label{sec:method_overview}
We adopt an embedding-based approach in which both entity mentions and type queries are embedded into a shared semantic space, enabling similarity-based search over their vector representations.

A central design choice is \textit{what} to embed and \textit{how} to embed it effectively. While previous approaches have relied on sentence-level representations \cite{rassin-etal-2024-evaluating} or prompted classification \cite{ashok2023promptner}, we argue that directly embedding individual entity mentions offers greater precision and flexibility. This approach allows the system to focus on localized signals, rather than diluting entity-type cues across entire documents. Although embedding every entity increases the number of stored vectors, we offset this by using low-dimensional representations tailored for named entity retrieval.

To obtain high-quality embeddings, we build on the contextual representations generated by LLMs. However, unlike most prior work that uses top-layer outputs, we systematically investigate which internal layer and subcomponent best captures entity-type information. We find that the value (V) vectors from a mid-layer (e.g., block 17 in LLaMA 3.1 8B) outperform other representations in distinguishing fine-grained types.

Using this insight, we construct an end-to-end system (\autoref{fig:main_architecture}) where both entity mentions and type queries are embedded using the selected LLM representation, followed by a lightweight multilayer perceptron (MLP) that maps these embeddings into a compact, type-aware vector space. The MLP is trained with a contrastive objective to align entities of the same type and separate those of different types. At indexing time, we compute and store the transformed embeddings for all entity mentions. At query time, we embed the user-specified type and retrieve the most similar entity vectors via nearest-neighbor search.

The next section provides a detailed analysis of LLM layer selection and motivates our choice of representation through a targeted evaluation of type-discriminative power.

\section{Contextual Entity Representations}
\label{sec:contextual_representation}
Our approach relies on extracting entity mention embeddings from intermediate layers of a pretrained LLM. In this section, we investigate which layer and component yield the most type-sensitive representations, that is, those that best distinguish entities based on their semantic class (e.g., \textit{drug} vs. \textit{disease}). While LLMs encode rich contextual information, not all layers capture type semantics equally well. Rather than using final-layer outputs, as is common in sentence embedding models, we empirically search for internal representations that maximize type separability.

\paragraph{Motivation: Embedding choice matters.} %
Prior work has shown that sentence embeddings, even those trained for retrieval, often underperform on entity-type discrimination tasks \citep{katz-etal-2023-neretrieve}. This occurs despite LLMs containing detailed entity knowledge \citep{heinzerling-inui-2021-language}, suggesting that naive pooling or top-layer representations fail to transfer type-specific information. We hypothesize that hidden representations from earlier layers may better isolate entity type information, and that these representations can be exploited for retrieval if appropriately selected.

\paragraph{Evaluation setup.}
We use the Few-NERD dataset to evaluate type sensitivity across different LLM layers and components. For each of 20 sampled fine-grained entity types, we collect 20 sentences containing an entity mention of that type. We then form balanced pairs: \textit{positive} pairs consist of two mentions of the same type, while \textit{negative} pairs consist of mentions of different types. For each mention, we extract its final-token representation and compute the cosine similarity between mention pairs. We then calculate AUC scores for a binary classification task (same-type or not).

\paragraph{Layer and component sweep.}
We apply this evaluation procedure to several LLMs, including LLaMA 3.1 8B, T5 11B, Mistral 7B, and Gemma 2 7B. For each model, we extract hidden representations from all transformer blocks and subcomponents (e.g., self-attention keys, values, output projections). In total, we analyze 416 distinct representation sources from LLaMA 3.1 alone. We repeat the full evaluation five times with different random seeds and report average AUC scores.

\begin{figure}[!ht]
    \centering
    \includegraphics[width=1\linewidth]{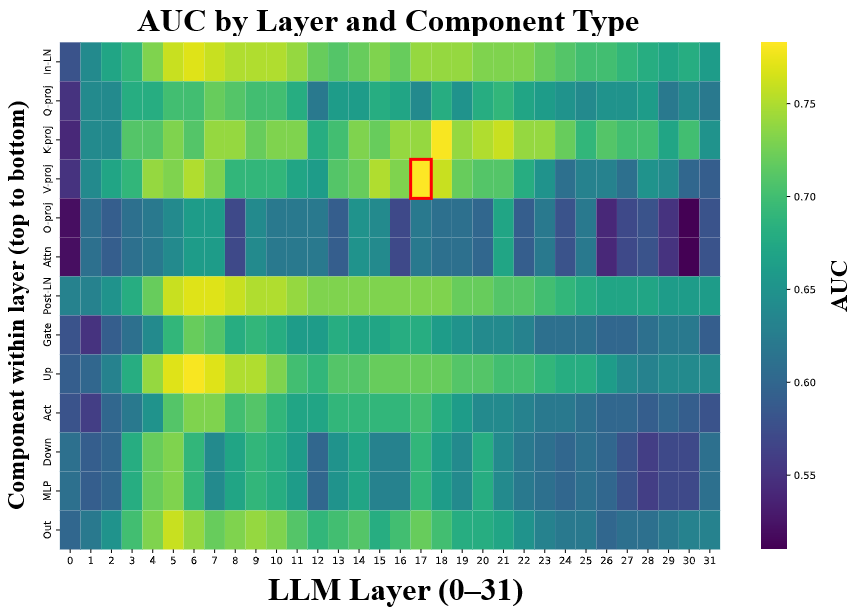}
    \caption{Entity-type discrimination AUC scores for 13 subcomponents across all 32 transformer blocks of LLaMA 3.1 8B.}
    \label{fig:llama-3-1-layers}
\end{figure}

\paragraph{Findings.}
\autoref{fig:llama-3-1-layers} shows a heatmap of AUC scores across LLaMA 3.1 layers and subcomponents. We observe substantial variation in performance, with mid-to-late layers outperforming both early and final layers. The highest AUC is achieved by the \textbf{value (V)} vectors in the self-attention module of block 17, suggesting that this component encodes particularly strong type-level signals.

\autoref{tab:top_performing_layers} compares top-performing layers across models. LLaMA 3.1 consistently yields better type sensitivity than other architectures, and the performance gap highlights the importance of both model choice and layer selection. 
Interestingly, the highest overall performance is specifically attributed to representations extracted from the value (V) vectors within the self-attention sub-layer of the 17th transformer block of Llama 3.1 8B, underscoring the significance of the self-attention mechanism in encoding detailed entity-type distinctions.
Moreover, it is clear that mid and lower layers encode entity type distinctions much better than top layers.

\begin{table}[!ht]
\centering
\resizebox{0.5\textwidth}{!}{
\begin{tabular}{|l|c|l|c|c|}
\hline
\textbf{Model} & \textbf{Block} & \textbf{Layer} & \textbf{Size} & \textbf{AUC} \\
\hline
Llama 3.1 8B & 17 & V & 1024 & 0.78 \\
T5 11B Encoder Part & 14 & Input Norm & 1024 & 0.74 \\
Gemma 2 7B & 6 & V & 4096 & 0.56 \\
Mistral 7B v0.3 & 9 & K & 1024 & 0.60 \\
\hline
\end{tabular}
}
\caption{Top-performing layer and subcomponent for each tested LLM.}
\label{tab:top_performing_layers}
\end{table}

\begin{figure}[!ht]
    \centering
    \includegraphics[width=1\linewidth]{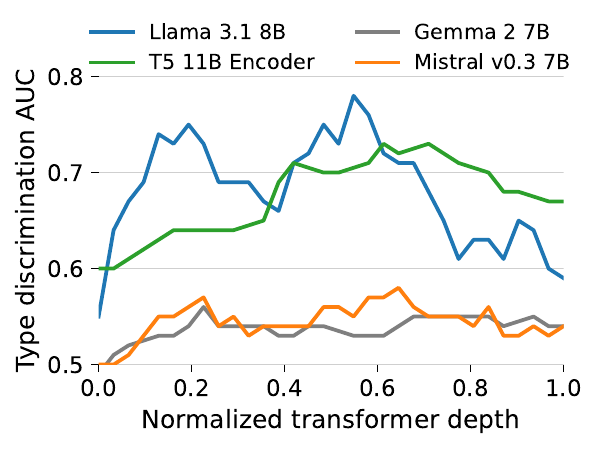}
    \caption{Type discrimination AUC for self-attention value (V) vectors across normalized transformer depth.}
    \label{fig:v-proj-all}
\end{figure}

\autoref{fig:v-proj-all} extends this analysis across LLMs by comparing value (V) vectors representations across all transformer blocks. To enable comparison across models with different depths, we normalize the block index from 0 (first block) to 1 (last block). This reveals that the trend persists across models: mid-to-late layers tend to better encode entity-type information, offering practical guidance for representation selection in entity-aware tasks.

Based on these findings, we use the value (V) vectors from block 17 of LLaMA 3.1 8B as the basis for entity embeddings in our retrieval system, described in the next section.

\section{Type-Aware Entity Embeddings}
\label{sec:type-aware}
LLM-derived contextual embeddings are high-dimensional and not inherently optimized for similarity-based named entity retrieval \citep{reimers-gurevych-2019-sentence}. To address this, we introduce a lightweight, supervised contrastive projection module that transforms entity representations into a compact, type-aware embedding space suitable for nearest-neighbor search (Type-Aware Entity Embeddings block in \autoref{fig:main_architecture}).

We use a two-layer multilayer perceptron (MLP) trained with a triplet contrastive loss \citep{Schroff_2015_CVPR}. 
Each training triplet includes three components: (1) an \textit{anchor}, which is a natural language type description (e.g., \textit{politician}); (2) a set of \textit{positives}, which are entity mentions of the same type (e.g., \textit{Angela Merkel}, \textit{George W. Bush}); and (3) a set of \textit{negatives}, which are mentions from different types (e.g., \textit{Danube River}, \textit{Mount Everest}).

Positive examples are sampled uniformly from entities sharing the same type, while negatives include both randomly selected and hard-negative entities from different types that exhibit lexical similarity to the anchor (e.g., sharing similar surface forms or contextual phrases). This encourages the model to distinguish semantically relevant types from superficially similar distractors.

The contrastive objective encourages embeddings of same-type entities to be close together and dissimilar types to be well separated. The resulting representations are low-dimensional, discriminative, and aligned with cosine similarity metrics used during retrieval \citep{weng2021contrastive}.

\subsection{Training Data}  
\label{Training Data}

We train our projection module using the NERetrieve dataset \citep{katz-etal-2023-neretrieve}, which includes 500 fine-grained entity types from diverse domains, each associated with thousands of weakly labeled paragraphs. We use the official training split, containing 400 entity types leaving the test entity types as queries for the end-to-end system evaluation. For each type, we generate 5,000 triplets comprising a query type, a positive paragraph (mentioning a relevant entity), and a negative paragraph (mentioning an unrelated entity). Negative paragraphs are sampled using the hard negative mining procedure described in \S\ref{lbl:impl_detil}. The full training set consists of 2 million triplets. 

\begin{figure*}[!ht]
  \centering
  \includegraphics[width=15cm,height=8cm]{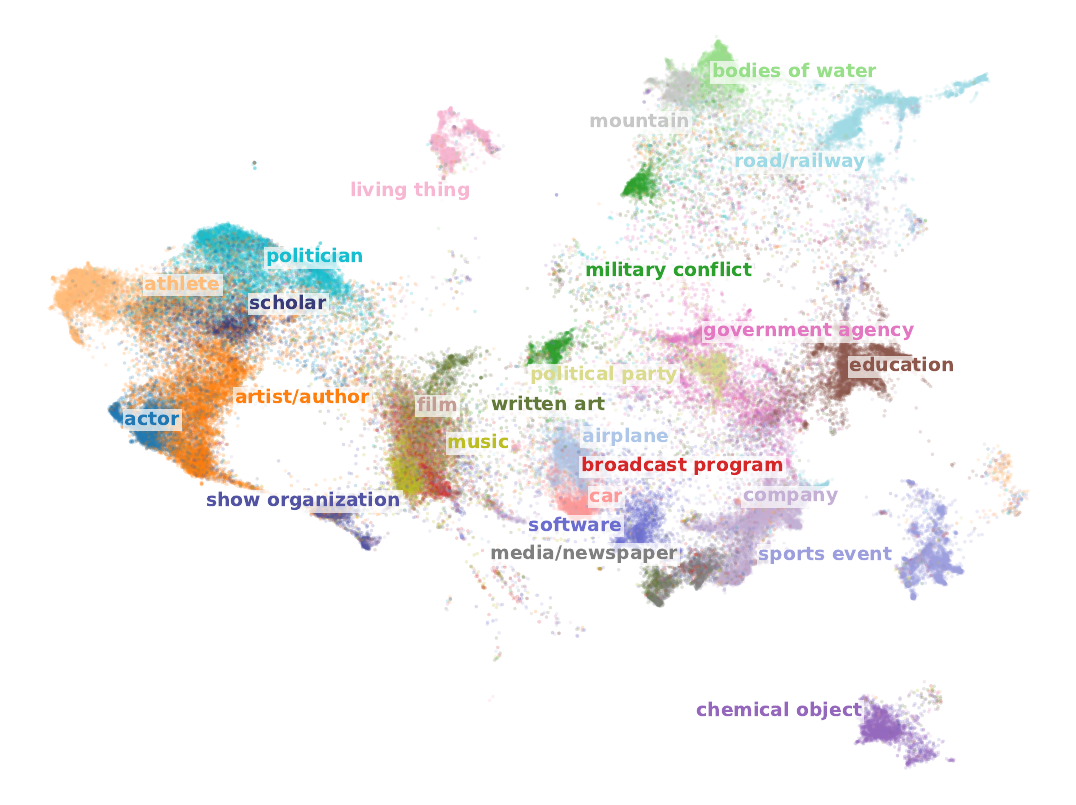}
  \caption{2D UMAP projection of type-aware entity embeddings produced by our model, visualized for the top 25 Few-NERD types. Entity mentions of the same type cluster tightly, while different types form distinct groups.}
  \label{fig:umap_scatter}
\end{figure*}

\subsection{Implementation details}
\label{lbl:impl_detil}
We use a frozen LLM (LLaMA 3.1 8B) as the base encoder and extract the final token’s value (V) vector from block 17 as the entity representation (see \S\ref{sec:contextual_representation}). This representation is passed through a two-layer MLP. 
Input layer's size is derived from the LLM output (in our case 1024). The output layer dimension was set to 500 to reduce the input size while preserving sufficient capacity for capturing task-specific information. The hidden layer was chosen to match this output dimension. We employ the SiLU activation function, which forms the core of SwiGLU used in many modern transformer LLMs, but without its additional parameters. This allows the projection network to learn effectively while remaining lightweight. We apply dropout with a small rate of 0.1 to mitigate overfitting while distilling the entity representations. We adopt Triplet Loss, where the anchor is the embedding of an entity type, positives are embeddings of entity spans of that type within their sentence context, and negatives are embeddings of entity spans belonging to other types. The objective is to align entity types with their instances while separating them from instances of other types.

For hard negative mining, 10\% of negative examples in each batch are selected using BM25 similarity to the anchor. This increases training difficulty and encourages robustness to lexical distractors \cite{karpukhin-etal-2020-dense}. 
All numerical hyperparameters were chosen by independently evaluating a range of values and selecting those yielding the best performance on the entity type discrimination task (Described in Section~\ref{sec:contextual_representation}).

\autoref{fig:umap_scatter} illustrates the effectiveness of our learned embedding space. Using UMAP \cite{McInnes2018}, we project type-aware embeddings of entity mentions from the top twenty five Few-NERD fine-grained types into two dimensions. The resulting layout reveals both fine and coarse-grained semantic structure. Entity mentions of the same fine type such as \textit{politician} or \textit{mountain} form dense, coherent clusters, indicating that the model successfully aligns entities of the same category. Moreover, clusters corresponding to different fine types that share a common coarse type such as \textit{geographic entities} (mountain, road and body of water) or \textit{people} (politician, artist and actor) tend to appear adjacent in the space. This suggests that the embedding space not only distinguishes between specific types but also preserves higher-level semantic relationships, supporting both fine-grained retrieval and type generalization.

\section{Experimental setting}
\label{sec:Experimental_setting}
\subsection{NER Retriever: End-to-End System}
\label{sec:retriever-system}
To evaluate the effectiveness of our type-aware entity embeddings in a retrieval setting, we implemented a complete NER Retriever system consisting of three main stages: \textbf{entity detection}, \textbf{indexing}, and \textbf{retrieval}, following the design described in Section~\ref{sec:method_overview} and illustrated in \autoref{fig:main_architecture}.
\paragraph{Entity detection phase.} In our approach, it is essential to identify the spans of all entities in a text, independent of their semantic category, so that they can be indexed in the subsequent phase. To this end, we employ an LLM trained on general NER corpora to mark entity spans, following the approach of \citet{awasthy2020cascaded}. The identifier operates in a category-agnostic manner, aiming to capture all entities present in the document. None of the benchmarks used in our evaluation were used in training this model.

\paragraph{Indexing phase.} For each document, all entity mentions are identified and the document is passed through a frozen LLM encoder (LLaMA 3.1 8B). The contextual representations corresponding to the end span of each entity are extracted from the selected internal component, the value (V) vector from transformer block 17 (see Section~\ref{sec:contextual_representation}). These embeddings are then passed through a trained projection module, a lightweight two-layer MLP that maps the high-dimensional LLM embeddings into a compact, type-sensitive space (Section~\ref{sec:type-aware}). The resulting entity vectors are stored in a vector index linked to their corresponding document and optimized for cosine similarity search.

\paragraph{Retrieval phase.} At query time, the user provides a type description in natural language (e.g., \dq{dinosaur} or \dq{airline}). This query is embedded using the same LLM and projection pipeline, producing a vector in the same embedding space as the indexed entities. We then retrieve the $k$ most similar entity embeddings based on cosine similarity, and return the corresponding documents as retrieval results. 
If a document contains multiple entities (and thus multiple embeddings), it is retrieved once as long as at least one of its embeddings meets the similarity criterion.

This setup allows us to evaluate how well our learned embedding space aligns entity mentions with arbitrary type queries in a zero-shot setting. The next subsections describe the datasets, evaluation protocol, and baselines used in our experiments.

\subsection{Baselines}
\label{baselines}
We evaluate our proposed NER Retriever framework against a set of strong baselines commonly used in representation-based retrieval tasks.

\paragraph{BM25.} A lexical retrieval baseline \citep{10.1561/1500000019}, BM25 remains competitive in many entity-centric retrieval settings \citep{ sciavolino2021simple}. We apply BM25 at the sentence level using the entity type label as a query and retrieve sentences containing matching entities.

\paragraph{E5-Mistral.} A decoder-only sentence embedding model that uses last-token pooling to produce fixed-size representations \citep{wang2023improving}. E5-Mistral is trained for general-purpose retrieval and serves as a strong dense baseline in both document and entity-level retrieval settings.

\paragraph{NV-Embed v2.} A recent decoder-only model with the same backbone as E5-Mistral but using a latent attention-based pooling mechanism to improve sentence-level representation quality \citep{lee2024nv}. \\

\noindent Both E5-Mistral and NV-Embed v2 have comparable architecture and parameter count to LLaMA 3.1, providing a fair basis for comparison with our approach.

All models are evaluated in the same zero-shot setting, without task-specific fine-tuning or instructions.

\begin{table*}[!t]
\centering
\begin{tabular}{|l|c|c|c|}
\hline
\textbf{Model} & \textbf{Few-NERD} & \textbf{MultiCoNER 2} & \textbf{NERetrieve Test} \\
\hline
    BM25                  & 0.22 & 0.08 & 0.27 \\
    NV-Embed v2    & 0.04 & 0.07 & \textbf{0.29} \\
    E5-Mistral            & 0.08 & 0.09 & 0.22 \\
    NER Retriever (Ours)     & \textbf{0.34$^{\dagger}$} & \textbf{0.32$^{\dagger}$} & 0.28 \\
\hline
    NER Retriever w/ \textit{entity span oracle}    &   0.37 & 0.35 & 0.34 \\
\hline
\end{tabular}
\caption{Zero-shot \textbf{R-Precision} scores across three datasets. NER Retriever yields up to four times more performance compared to lexical (BM25) and dense (E5-Mistral, NV-Embed v2) baselines. $\dagger$ indicates that the score is significantly higher than the strongest competing baseline (Wilcoxon, $p<0.05$)}
\label{tab:model_results}
\end{table*}

\subsection{Evaluation Datasets} 
\label{sec:evaluation_datasets}
We evaluate our system in a zero-shot entity retrieval setting using multiple fine-grained NER benchmarks adapted as retrieval tasks. In each benchmark, every entity type is treated as a distinct query, and the goal is to retrieve all text segments (paragraphs or sentences) containing entity mentions of that type.

We conduct experiments on three widely used NER benchmarks, repurposed for entity-centric retrieval:

\paragraph{NERetrieve} \citep{katz-etal-2023-neretrieve}: The NERetrieve test set is a large, silver-annotated benchmark comprising 100 held-out fine-grained entity types paired with approximately 2.4 M paragraphs. We use the test split described in \S\ref{Training Data}, which is disjoint from the contrastive MLP training data. Due to the substantial size of this split, evaluation is performed on a randomly sampled 5\% subset, corresponding to about 120 K documents, to ensure computational feasibility while maintaining representativeness.

\paragraph{Few-NERD (supervised)} \citep{ding-etal-2021-nerd}: A manually annotated dataset covering 66 fine-grained entity types across 188K Wikipedia sentences. For fairness in comparison, we omit instances with fewer than five words, as such very short sentences were found to harm the performance of sentence-encoder baselines while not impacting our method.
    
\paragraph{MultiCoNER 2} \citep{fetahu-etal-2023-multiconer}: A silver-annotated dataset containing 12 languages and 33 entity types. We used the English section in evaluation, consisting of \(\sim\)268K sentences. Extracted from multiple sources, it emphasizes short texts and low-context scenarios, making it a challenging benchmark for retrieval-based systems.

These benchmarks include long-tail and domain-specific types, making them well-suited to evaluate zero-shot generalization. Crucially, all test-time entity types and associated documents are disjoint from those used during training our system.

\subsection{Evaluation Metrics} 
We report \textbf{Average R-Precision} \citep{Manning_Raghavan_Schütze_2008} to assess retrieval performance. For each query (i.e., entity type), R-Precision is defined as Precision@K, where $K$ is the total number of relevant documents for that query. We compute the R-Precision for each entity type and report the mean across all queries in a dataset. This metric rewards systems that retrieve all relevant mentions while penalizing retrieval of unrelated entities.

\section{Results and Analysis}

\label{sec:results}

\autoref{tab:model_results} reports R-Precision for NER Retriever and baseline models across three benchmarks (Few-NERD, MultiCoNER 2, and NERetrieve). NER Retriever substantially outperforms all baselines on Few-NERD and MultiCoNER 2, while achieving comparable performance on the NERetrieve test set, demonstrating the effectiveness of type-aware embeddings for ad-hoc entity retrieval.

\paragraph{MultiCoNER 2.} This dataset presents the most challenging retrieval scenario due to its short and low-context text segments. NER Retriever achieves an R-Precision of 0.32—more than three times higher than E5-Mistral (0.09) and four times higher than BM25 (0.08). These results indicate that sentence-level embedding models and lexical methods struggle in limited-context settings, while our type-aware embeddings remain effective.

\paragraph{Few-NERD.} NER Retriever also leads on Few-NERD with an R-Precision of 0.34, substantially outperforming both NV-Embed v2 (0.04) and E5-Mistral (0.08). This confirms our system’s ability to generalize across manually annotated, diverse fine-grained entity types.

\paragraph{NERetrieve Test.} On the large and diverse NERetrieve Test set, NER Retriever achieves results comparable to NV-Embed v2 (0.29 vs. 0.28, difference not statistically significant) and BM25 (0.28). This is the only dataset where our method does not clearly outperform all baselines. We attribute this to the dataset’s descriptive nature, as its Wikipedia-based source often provides explicit cues about entity types. For example, when querying for “French singer,” the term singer is frequently stated directly in the text. The fact that BM25 also performs competitively supports this interpretation.

\paragraph{Entity detection.} \autoref{tab:model_results} highlights the critical role of entity span detection in our framework. When we replace the automatic span detector with an oracle that uses the gold annotations provided by each dataset, NER Retriever’s performance improves by roughly 11\% on average. This gap illustrates how missed or incorrectly detected spans can directly limit retrieval coverage. Importantly, under the oracle setting, NER Retriever outperforms all baselines, suggesting that our retrieval approach is highly effective when provided with accurate spans. These results underscore that advances in entity detection can further enhance NER Retriever’s robustness and overall effectiveness.

\begin{table*}[!ht]
\centering
\begin{tabular}{lccc}
  \toprule
  \textbf{Model} & \textbf{Few\text{-}NERD} & \textbf{MultiCoNER 2} & \textbf{NERetrieve Test} \\
  \midrule
  BM25                       & 0.42/0.38 & 0.29/0.27 & 0.67/0.52 \\
  E5-Mistral                         & 0.33/0.10 & 0.34/0.27 & 0.42/0.42 \\
  Nvidia NV\text{-}Embed     & 0.03/0.02 & 0.28/0.18 & 0.35/0.33 \\
  NER Retriever  & 0.49/0.48 & 0.39/0.42 & 0.49/0.41 \\
  \hline
  NER Retriever w/ \textit{entity span oracle}     & 0.61/0.58 & 0.50/0.49 & 0.61/0.52 \\
  \bottomrule
\end{tabular}
\caption{\textbf{Precision@50 / Precision@200} for different models across datasets}
\label{tab:precision_combined}
\end{table*}
We report additional metrics (Precision@50 and Precision@200) in Table~\ref{tab:precision_combined}, which demonstrate NER Retriever’s effectiveness even in non-exhaustive retrieval scenarios, maintaining a strong lead in top-k results.

\paragraph{Statistical significance.} 
We tested significance using the Wilcoxon signed-rank test. For each dataset, we compared our method against the strongest alternative (i.e., the highest-scoring competing model) to assess whether any other system significantly outperforms ours. All significance claims are based on a $p$-value threshold of 0.05.

\section{Ablation Studies and Analysis}

We conduct a series of ablation experiments to isolate the contribution of key design choices in our pipeline: (i) selection of the hidden layer used for representation extraction; (ii) method for selecting the token representation of an entity; and (iii) use of a task-specific projection MLP. All ablation experiments were evaluated on Few-NERD.

\subsection{Hidden Layer Selection}
Our approach uses the V-projection output from transformer block 17 of the LLaMA 3.1 8B  model to extract contextual representations. Exhaustive evaluation of all possible layers is computationally infeasible. Instead, we adopted a layer estimation method (see Section~\ref{sec:contextual_representation}) to identify a layer likely to contain task-relevant information. To validate this choice, we compared retrieval performance when using representations from the selected hidden layer versus the final model output. Results show a significant improvement in average R-Precision from 0.09 (final layer) to 0.19 (block 17), demonstrating the effectiveness of our layer selection strategy.
\subsection{Token Representation}

We compare two strategies for selecting the token embedding used to represent each entity. The first uses the representation of the end-of-sequence (EOS) token, hypothesizing that it may capture a global summary of the sentence. The second uses the final token in the span corresponding to the entity, motivated by the decoder’s autoregressive property, only the final token can attend to all previous tokens in the sequence. As shown in Table~\ref{tab:ablation_table}, span-based representations substantially outperform EOS-based ones (0.19 vs. 0.03 R-Precision), supporting the use of direct entity-span tokens for retrieval tasks.

\subsection{MLP Projection}

We assess the effect of applying a learned MLP on top of the contextual embedding. The MLP serves to both distill the high-dimensional LLM representation into a lower-dimensional, task-specific embedding space and improve retrieval performance. We also experiment with feeding both the EOS token and the entity token as a dual input to the MLP, but this configuration yields no performance gain over using only the entity token, as shown in Table~\ref{tab:ablation_table}.

\begin{table}[!ht]
\centering
\small
\begin{tabular}{|l|c|}
\hline
\textbf{Model Variant} & \textbf{R-Precision} \\
\hline
Full (17V + Entity  MLP)  & \textbf{0.34} \\
17V + Entity \& EOS MLP & 0.33 \\
17V + Entity (-No MLP) & 0.16 \\
Final Output Layer + Entity (-No MLP) & 0.08 \\
17V + EOS (-No MLP) & 0.02 \\
\hline
\end{tabular}
\caption{Ablation results on the Few-NERD dataset, measuring the effect of different components in the NER Retriever architecture.}
\label{tab:ablation_table}
\end{table}

\subsection{Efficiency and Storage Considerations}
NER Retriever introduces a shift in retrieval design by storing one embedding per entity instance rather than per sentence. While this increases the number of stored vectors, each vector is only 500 dimensions, substantially smaller than modern sentence embedders like NV-Embed v2 (4096 dimensions). Moreover, NER Retriever does not generate embeddings for text segments without detected entities, further reducing storage requirements and potentially improving retrieval efficiency compared to sentence-level models that embed every document regardless of content. On MultiCoNER 2, the dense vector index for NV-Embed v2 occupies 9.2 GB, whereas NER Retriever requires only 2 GB, a reduction of nearly 79\%.

\section{Related Work}

Named Entity Recognition (NER) has evolved from identifying coarse-grained types such as PERSON, ORGANIZATION, and LOCATION \citep{tjong-kim-sang-de-meulder-2003-introduction,settles2004biomedical,sekine-2008-extended} toward fine-grained and domain-specific categories. Fine-grained entity typing seeks to assign open-ended, context-sensitive type labels, with \citet{choi-etal-2018-ultra} introducing ultra-fine typing using natural language phrases.  

Large language models (LLMs) have recently been shown to perform zero-shot NER by leveraging their parametric world knowledge \citep{ashok2023promptner,xie-etal-2023-empirical}. Our approach builds on this intrinsic capability, extracting entity representations from LLMs to enable zero-shot retrieval of named entities.  

Named Entity Retrieval, introduced by \citet{katz-etal-2023-neretrieve}, frames the task of retrieving all documents containing mentions of entities of a given type. Existing retrieval methods perform poorly on this task, and entity-focused benchmarks such as QUEST \citep{malaviya-etal-2023-quest}, DBpedia Entity v2 ListSearch \citep{Hasibi:2017:DVT}, QAMPARI \citep{amouyal-etal-2023-qampari}, and D-MERIT \citep{rassin-etal-2024-evaluating} similarly show that off-the-shelf retrieval achieves limited effectiveness. Prior work has focused mainly on linking entity mentions to knowledge bases \citep{de2020autoregressive,wu-etal-2020-scalable} or supervised entity retrieval in QA \citep{sciavolino2021simple,shavarani-sarkar-2025-entity}, but none explicitly address ad-hoc retrieval of free-form entity categories in text—a gap our work targets.  

As shown in the WideSearch benchmark \citep{wong2025widesearch}, agents are limited in handling broad queries that demand exhaustive coverage; NER Retriever advances this goal by offering fine-grained named entity retrieval that could equip RAG-based agents with the granular evidence needed for reliable broad-query search.  

Our method employs contrastive learning, aligning entity-category queries with entity-mention spans. While contrastive learning has been explored for traditional NER \citep{das-etal-2022-container,zhangoptimizing}, these works focus on token classification rather than retrieval-based scenarios.
\section{Conclusion}
We introduced NER Retriever, a zero-shot framework for ad-hoc named entity retrieval based on type-aware embeddings derived from LLMs. By selecting and refining mid-layer representations with contrastive learning, our method enables accurate and efficient retrieval across diverse, unseen entity types. Our results highlight the effectiveness of representation selection and suggest new directions for embedding-based NER systems.

\section*{Limitations}
Our system relies on the parametric knowledge encoded in LLMs, which may limit its effectiveness in specialized domains such as law, medicine, or finance. In zero-shot settings, performance in these areas may degrade due to the lack of domain-specific coverage. This limitation could be addressed by integrating domain-adapted LLMs or fine-tuning with targeted data.

\bibliography{custom}

\clearpage
\appendix

\section{System Implementation}

\subsection{Entity Detection}
In order to identify entities in text, we utilized the extractor component of CascadeNER~\cite{awasthy2020cascaded}\footnote{\url{https://huggingface.co/CascadeNER/models_for_CascadeNER}}, a Qwen~2.5 model fine-tuned for entity span detection. The model was trained on the DynamicNER dataset, which provides multilingual and fine-grained entity annotations; importantly, none of the evaluation datasets used in this work were included in its fine-tuning. For each input text, the extractor outputs a marked version in which entities are encompassed by \#\# symbols. For example:

\noindent\texttt{model output: "\#\#Claremore Lake\#\# is a reservoir in \#\#Rogers County\#\#"}

We evaluate the entity detection phase using the original span annotations provided in each dataset. A match between two spans is considered valid if their character-level Jaccard similarity exceeds 0.8. We report the coverage of annotated entities for each evaluation dataset, while noting that the extractor also identifies many additional entities that fall outside the original dataset's annotation scope. As a result, the extractor enables a more exhaustive identification process than what is captured by the evaluation datasets. Overall, the results are very strong, though there remains room for improvement.

\begin{table}[H] %
\centering
\caption{Entity detection coverage using the entity extractor model}
\label{tab:entity_recall}
\begin{tabular}{l S[table-format=1.2]}
\toprule
\textbf{Dataset} & \multicolumn{1}{c}{\textbf{Coverage}} \\
\midrule
NERetrieve    & 0.89 \\
MultiCoNER~2 & 0.90 \\
Few\text{-}NERD & 0.94 \\
\bottomrule
\end{tabular}
\end{table}

\end{document}